# Spin-orbit torque as a method for field-free detection of in-plane magnetization switching

Nguyen Huynh Duy Khang[1,2] and Pham Nam Hai[1,3,4*]

[1]*Department of Electrical and Electronic Engineering, Tokyo Institute of Technology,*

*2-12-1 Ookayama, Meguro, Tokyo 152-8550, Japan*

[2]*Department of Physics, Ho Chi Minh City University of Education,*

*280 An Duong Vuong Street, District 5, Ho Chi Minh City 738242, Vietnam*

[3]*Center for Spintronics Research Network (CSRN), The University of Tokyo,*

*7-3-1 Hongo, Bunkyo, Tokyo 113-8656, Japan*

[4]*CREST, Japan Science and Technology Agency,*

*4-1-8 Honcho, Kawaguchi, Saitama 332-0012, Japan*

*Corresponding author: pham.n.ab@m.titech.ac.jp

**Abstract:** We proposed and demonstrated a simple method for detection of in-plane magnetization switching by the spin-orbit torque (SOT) in bilayers of non-magnetic / magnetic materials. In our method, SOT is used not only for magnetization switching but also for detection. Our method can detect arbitrary $M_x$ and $M_y$ component without an external magnetic field, which is useful for fast characterization of type-X, type-Y, and type-XY SOT magnetization switching. Our SOT detection scheme can be utilized not only for fast characterization of SOT switching in bilayers, but also for electrical detection of in-plane magnetic domains in race-track memory.

Spin-orbit torque (SOT) has increasingly attracted attention as an emerging method for magnetization switching in magnetoresistive random access memory (MRAM).[1,2] For fast evaluation and optimization of the SOT effect, bilayers of a non-magnetic spin Hall layer and a magnetic layer are usually fabricated and evaluated, without implementation of full stack three-terminal magnetic tunnel junctions.[3,4] Therefore, other methods rather than the tunneling magnetoresistance effect have to be used for detection of SOT-induced magnetization switching. If the magnetic layer has perpendicular magnetic anisotropy (PMA), i.e. in the so-called type-Z switching, the magnetization direction can be easily detected by the anomalous Hall effect (AHE). Although type-Z switching is attractive for potentially high bit-density and ultrafast (sub-ns) SOT-MRAM, the switching current density is as high as $10^7$ to $10^8$ Acm$^{-2}$, which would require very large Si transistors.[5] Furthermore, an in-plane external magnetic field is typically required for symmetry breaking. On the other hand, when the magnetization is in the film plane and perpendicular to the writing current direction (type-Y switching), the switching current density can be reduced to the order of $10^6$ Acm$^{-2}$ and no external magnetic field is required,[6] but with the expense of lower bit density and longer switching time. In this case, since the magnetization is in-plane, a differential planar Hall effect (PHE) technique was proposed for detection of magnetization switching, as shown in Fig. 1(a).[7,8] In the differential PHE technique, after each writing pulse current, an alternating magnetic field is applied to slightly tilt the magnetization toward the $x$ direction, which would generate a difference in the planar Hall voltage under a small reading current when the magnetization is reversed. In principle, this technique can also be used for detection of the type-X switching when the magnetization is in the film plane and aligned along the current direction.

In this work, we propose an alternative method for detection of in-plane SOT magnetization switching without an external magnetic field. In our method, SOT is used not only for magnetization switching but also for detection. Figure 1(b) illustrates our SOT detection scheme for type-X switching. After each writing pulse current, we apply a sinusoidal reading current along the $x$ direction. This reading current generates several alternating SOT effective fields, such as an antidampinglike effective field $H_{AD}$ that tilts the magnetization vector toward the $z$ direction, or a fieldlike effective field $H_{FL}$ that tilts the magnetization vector toward the $y$ direction. In Fig. 1(b), we show only the $H_{AD}$ component for sake of simplicity. Thus, there arise alternating $z(y)$ components of the magnetization that can be detected via AHE and PHE, respectively. There are also Hall signals from the anomalous Nernst effect (ANE) and the spin Seebeck effect (SSE) which are proportional to $\nabla T_z M$, where $\nabla T_z$ is the $z$ direction temperature gradient. The Hall signals from those effects can be detected with high signal-to-noise ratio by measuring the second harmonic responses with a lock-in amplifier. Our method does not require the external magnetic field for detection of magnetization switching as in the case of differential PHE technique. Note that our method can also be applied for type-Y switching. In this case, the SOT effect by the reading current along the $x$ direction is negligible, thereby one needs to apply the reading current along the $y$ direction and read the Hall voltage along the $x$ direction.

To demonstrate our SOT detection scheme, we prepared a stack consisting of Si/SiO$_2$ substrate / Fe (1 nm) / Pt (0.8 nm) / BiSb (10 nm) by magnetron sputtering, as illustrated in Fig. 2(a). Here, BiSb is a topological insulator with high electrical conductivity and giant spin Hall angles.[9,10,11,12] The very thin Pt layer is inserted between BiSb and Fe to prevent diffusion of Bi/Sb atoms to the Fe layer, and to add interfacial perpendicular magnetic anisotropy that reduces the in-plane magnetic anisotropy of the Fe layer. Figure 2(b) shows the magnetization curves of the Fe layer measured with in-plane and out-of-plane magnetic fields. The normalized saturation magnetization is 2105 emu/cc, which is higher than that of pure Fe due to the proximity-induced magnetic moment of Pt. We then used optical lithography and ion-milling to fabricate a Hall cross bar structure with width of 20 μm. Figure 2(c) shows an optical image of a Hall bar device and the experimental setup for type-X switching. Figure 2(d) shows the anomalous Hall resistance of the device measured with a perpendicular magnetic field and a constant current, which indicates an effective in-plane magnetic anisotropy field $H_k^{\text{eff}}$ of 4.3 kOe. Figures 2(e) and 2(f) show the first harmonic Hall resistance $R_\omega^{xy}$ and the second harmonic Hall resistance $R_{2\omega}^{xy}$ measured with an alternating current density $J^{\text{BiSb}} = 0.33\times10^5$ Acm$^{-2}$ at the frequency $\omega/2\pi$ of 259.68 Hz, under a magnetic field sweeping along the $x$ direction. $R_\omega^{xy}$ shows jumps corresponding to PHE. Meanwhile, $R_{2\omega}^{xy}$ shows clear hysteresis corresponding to the magnetization reversal by the in-plane magnetic field $H_x$. This indicates that it is possible to detect magnetization switching along the $x$-axis by the SOT effect from a small reading current.

In Fig. 3(a), we describe details of type-X magnetization switching and detection by SOT. When the reading current is applied along the $x$ direction, because $\boldsymbol{H}_{\text{AD}} \propto \widehat{\boldsymbol{\sigma}} \times \widehat{\boldsymbol{m}}$ where $\widehat{\boldsymbol{\sigma}}$ is the spin polarization unit vector of the spin current along the $y$ direction and $\widehat{\boldsymbol{m}}$ is a unit vector of the magnetization direction, $H_{\text{AD}}$ is maximum when $\boldsymbol{M}$ is aligned along the $x$ axis. However, If $\boldsymbol{M}$ is misaligned from the $x$-direction by a small angle $\phi$, there are both $M_x = M\cos\phi$ and $M_y = M\sin\phi$ components. The $M_x$ and $M_y$ components can be detected with the reading current applied along the $x$ and $y$ direction, respectively. Figure 3(b) shows representative SOT switching loops for type-X switching of our device with the reading current applied along the $x$ and $y$ direction. Before each measurement, we applied a large magnetic field to the sample along the $x$ direction, then reduced the magnetic field to zero. We subsequently applied a train of 100 ms writing pulse currents with gradually changing magnitude along the $x$ direction for SOT switching. Between each pulse, we detected the magnetization direction following the scheme in Fig. 3(a) with a reading current of $J^{\text{BiSb}} = 0.33\times10^5$ Acm$^{-2}$ applied along the $x$ and $y$ direction for detection of the $M_x$ and $M_y$ component. If *M* is perfectly aligned along the *x* direction, an external perpendicular magnetic field $H_z$ is needed to break the symmetry for type-X switching, and the switching direction depends on the direction of the external magnetic field.[2] However, we found that our type-X switching happens even at zero magnetic field. Indeed, SOT switching at zero magnetic field is nearly the same as that at $H_z$ = 170 Oe, with the small critical switching current $J_C^{BiSb}$ of 4.5×10$^5$ Acm$^{-2}$ thanks to the giant spin Hall angle of BiSb. Furthermore, applying a negative $H_z$ = -170 Oe slightly increased the switching current density to 5.5×10$^5$ Acm$^{-2}$ but did not change the switching direction. We note that the zero-field type-X SOT switching and the insensitivity to perpendicular magnetic fields were already observed by Fukami *et al.*[13] in elliptical SOT devices whose easy axis was slightly misaligned from the *x* direction by using shape magnetic anisotropy, referred here as type-XY. In our device, the misalignment is evidenced from the existence of the $M_y$ component detected with the reading current applied along the *y* direction. Using $\tan\phi = R_{2\omega}^{xy}(J_{read}//y)/R_{2\omega}^{xy}(J_{read}//x)$, the misalignment angle is estimated to be about 13 degree. Note that the field-free switching is very reproducible, as we observed sequential multiple switching up to at least 100 cycles as shown in Fig. 3(c).

Next, we discuss about the origin of such misalignment. First, there is a possibility that magnetization is accidentally misaligned from the *x* direction during initialization by the large in-plane magnetic field we applied along the *x* direction before each measurement. However, such accidental misalignment would be gradually randomized after multiple switching and thus cannot explain the very reproducible switching in Fig. 3(c). Thus, an intrinsic easy axis is needed for field-free switching. Such an easy axis can originate from shape magnetic anisotropy, crystalline magnetic anisotropy, or strain-induced magnetic anisotropy during sputtering deposition. Given the fact that our device is a symmetric Hall cross bar and the Fe layer is polycrystalline, we suggest that the strain-induced magnetic anisotropy during the sputtering process is likely the origin of the easy axis. Further works will be needed to clarify details of the observed field-free switching.

Finally, to quantitatively evaluate the SOT effect in our device, we measured $H_{AD}$ and the effective spin Hall angle $\theta_{SH}^{eff}$ by using the high field second harmonic technique. Figure 4(a) shows $R_{2\omega}^{xy}$ measured up to $H_x = \pm 8$ kOe at different $J^{BiSb}$. At high magnetic fields, $R_{2\omega}^{xy}$ is given by [14]

$$R_{2\omega}^{xy} = \frac{R_{AHE}}{2}\frac{H_{AD}}{H_x+H_k^{eff}} + R_{PHE}\frac{H_{FL}}{H_x} + R_{ANE+SSE} \quad (1).$$

Here, $R_{AHE} = 6.4\ \Omega$, $R_{PHE} = 0.08\ \Omega$, and $H_k^{eff} = 4.3$ kOe. The first, second and third term represent contribution from the antidamplinglike effective field $H_{AD}$ via the AHE effect, the fieldlike effective field $H_{FL}$ via the PHE effect (which is very small since $R_{AHE} \gg R_{PHE}$ and $H_{AD} \gg H_{FL}$ for topological insulators),[10,15] and the thermal effect including ANE and SSE. By fitting the experimental data to equation (1), we can obtain $H_{AD}$, which are plotted against $J^{BiSb}$ in Fig. 4(b). From the $H_{AD}$ - $J^{BiSb}$ gradient, we obtain $\theta_{SH}^{eff} = 2.4$ for BiSb. Considering the spin loss in the 0.8 nm-thick Pt layer, the intrinsic spin Hall angle $\theta_{SH}$ of BiSb is estimated to be about 3.1, which is similar to that reported in (Co/Tb)$_n$/Pt/BiSb heterostructure,[11] but smaller than that in (Co/Pt)$_n$/BiSb heterostructure.[12] From these data, $H_{AD}$ at the critical switching current density $J^{BiSb} = 4.5\times10^5$ Acm$^{-2}$ is estimated to be about 17 Oe. This SOT field is close to the coercive force (depinning field) observed in Fig. 2(f), confirming that the SOT effect is the driving mechanism for the magnetization switching in Figs. 3(b) and 3(c).

In summary, we have proposed and demonstrated a simple method for detection of in-plane magnetization switching by SOT. Our method can detect the arbitrary $M_x$ and $M_y$ component without an external magnetic field, which is useful for fast characterization of type-X, type-Y, and type-XY SOT magnetization switching. Last but not least, our SOT detection scheme can be used for electrical detection of in-plane magnetic domains in race-track memory without using Kerr effect microscopy or magnetic force microscopy.

**Acknowledgment**

This work is supported by JST-CREST (JPMJCR18T5). N.H.D.K. acknowledges Marubun Research Promotion Foundation for an exchange research grant, and JSPS for a postdoctoral fellowship for research in Japan (P20050). The authors thanks S. Nakagawa Laboratory at the Tokyo Institute of Technology for the support and help of magnetization measurement.

**Data Availability Statement**

The data that support the findings of this study are available from the corresponding author upon reasonable request.

**Figures and Captions**

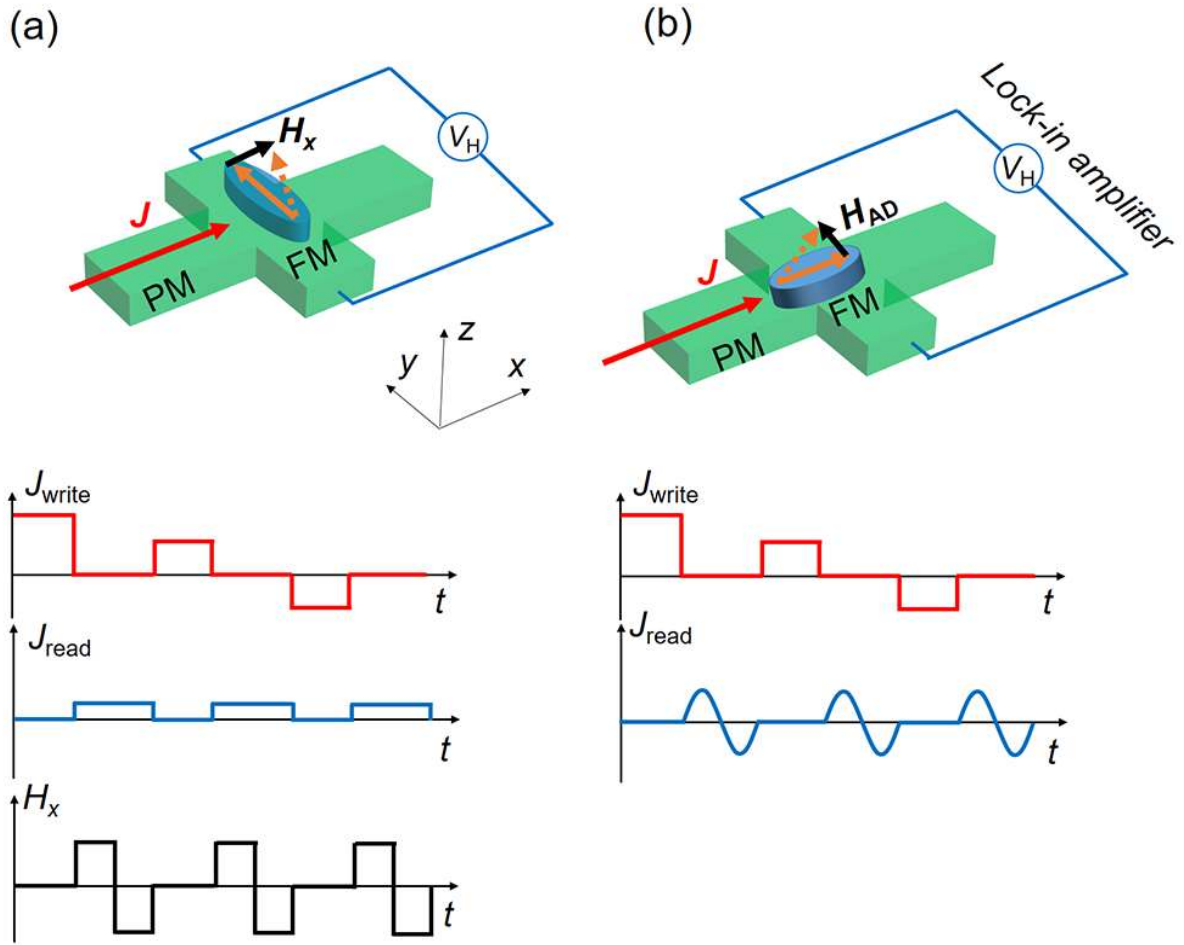

**Figure 1. (a)** Detection scheme for type-Y SOT switching using differential planar Hall effect. Alternating external magnetic field is applied to *x* direction during read out. **(b)** Our proposed SOT detection scheme for type-X SOT switching. No external magnetic field is required for read out.

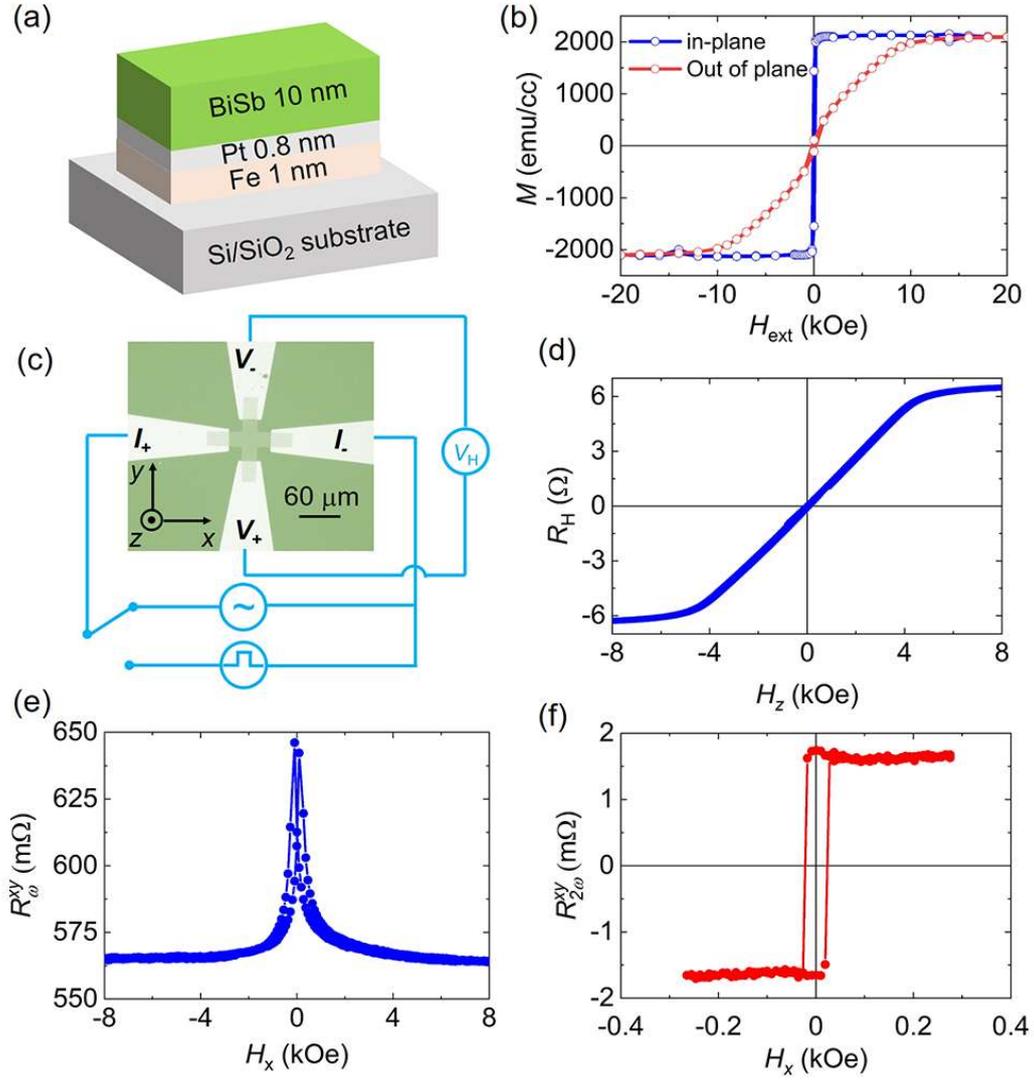

**Figure 2. (a)** Thin film stack for demonstration of proposed SOT detection scheme. **(b)** Magnetization curves of Fe layer measured with in-plane and out-of-plane magnetic fields. **(c)** Optical image of Hall bar device and experimental setup for type-X switching. **(d)** Anomalous Hall resistance of device measured with perpendicular magnetic field and constant current. **(e),(f)** first and second harmonic Hall resistance measured with alternating current density $J^{BiSb}$ = 0.33×10$^5$ Acm$^{-2}$ at frequency $\omega/2\pi$ of 259.68 Hz, under magnetic field sweeping along $x$ direction.

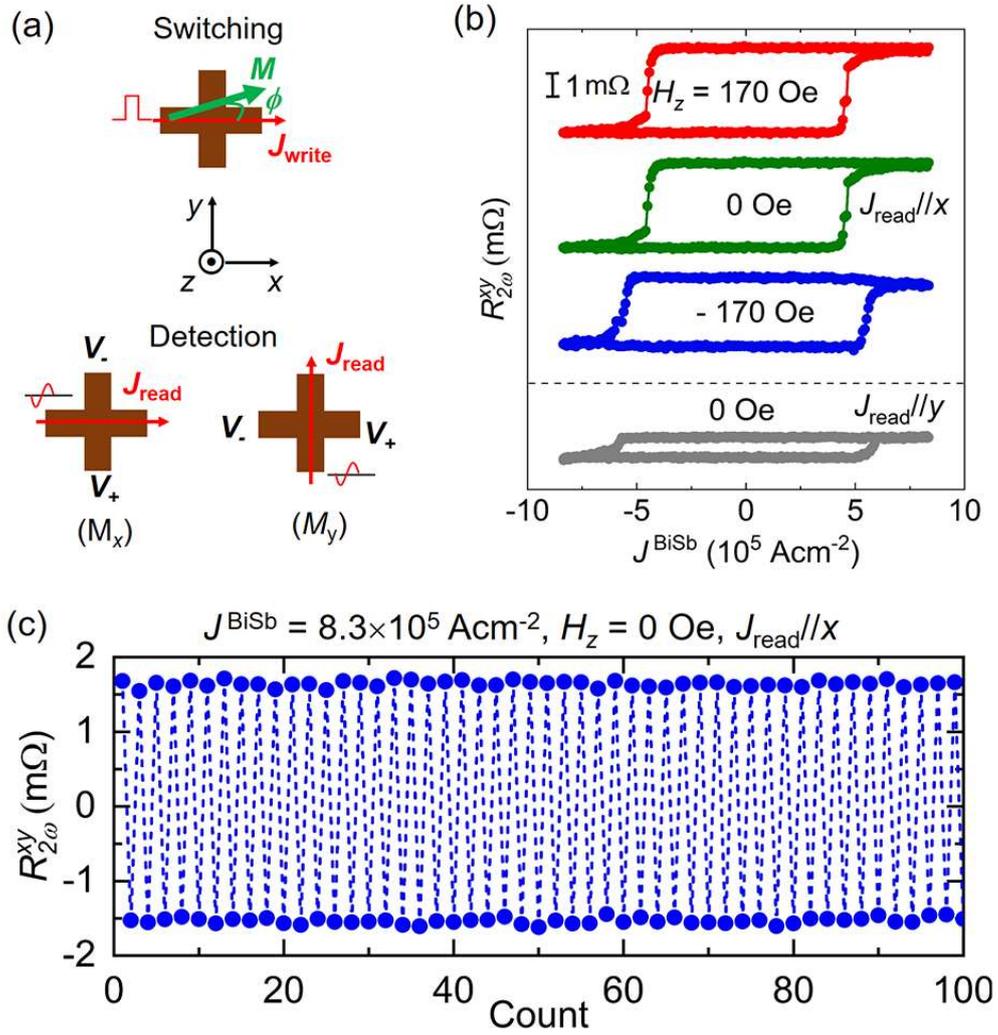

**Figure 3. (a)** Details of type-X SOT switching and SOT detection. In-plane $M_x$ and $M_y$ component are detected by applying reading current along $x$ and $y$ direction, respectively. **(b)** Representative SOT switching loops measured with reading current applied along $x$ direction (upper) and $y$ direction (lower). **(c)** Field-free sequential type-X SOT switching by $J^{BiSb}$ = $8.3 \times 10^5$ Acm$^{-2}$, detected by reading current applied along $x$ direction.

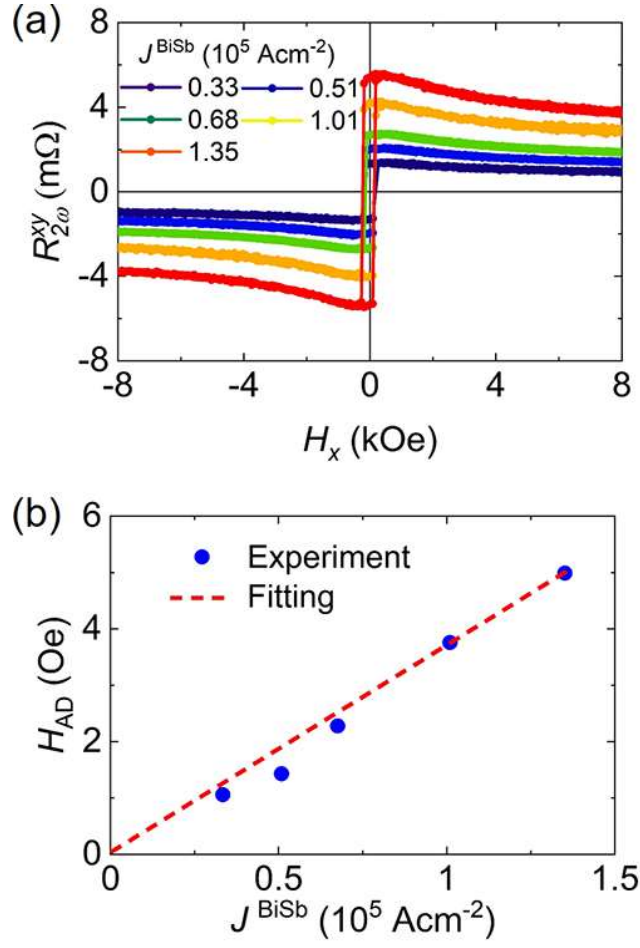

**Figure 4. (a)** High field second harmonic Hall resistance measured up to ±8 kOe at various $J^{BiSb}$. **(b)** $H_{AD}$ obtained by fitting experimental data in (a) to equation (1).